\documentclass[proceedings, twocolumn]{rmaa}
\usepackage{rmaacite}
\renewcommand{\P}[1]{%
\ifnum#1=1\hbox{OW~168--326E}\fi
\ifnum#1=2\hbox{OW~167--317}\fi
\ifnum#1=3\hbox{OW~163--317}\fi
\ifnum#1=5\hbox{OW~158--323}\fi
\ifnum#1=0\hbox{OW~171--334}\fi}

\title{The O II Recombination Line Abundance Problem in Planetary Nebulae}
\author{D. R. Garnett\altaffilmark{1} and H. L. Dinerstein\altaffilmark{2} }
\altaffiltext{1}{Steward Observatory, University of Arizona.}
\altaffiltext{2}{University of Texas at Austin, Austin.}
\fulladdresses{
\item D. R. Garnett: Steward Observatory, University of Arizona
  933 North Cherry Avenue, Tucson, AZ 85721 USA (dgarnett@as.arizona.edu)
\item H. L. Dinerstein: Astronomy Department, University of Texas at Austin,
  Austin, TX 78712 USA (harriet@astro.as.utexas.edu) }
\shortauthor{Garnett \& Dinerstein}
\shorttitle{O II Recombination Lines in Planetary Nebulae}

\abstract{  
We present new observations of O II recombination lines in ten 
bright planetary nebulae, along with spatially-resolved 
measurements of O II and [O~III] in the Ring nebula NGC 6720, to
study the discrepancy between abundances derived from 
O~II recombination lines and those derived from 
collisionally-excited [O~III]. We see a large range 
in the difference between O~II- and [O~III] derived abundances,
from no difference up to a factor six difference. The size of this
discrepancy is anti-correlated with nebular surface brightness; 
compact, high-surface-brightness nebulae have the smallest discrepancies. 
O~II levels that are populated mainly by dielectronic recombination 
give larger abundances than other levels. Finally, our long-slit 
observation of the Ring nebula shows that the O~II emission peaks 
$interior$ to the bright shell where [O~III] and H$\beta$ are 
strongest. 
Based on the observed correlations, we propose that the
strong recombination line emission in planetary nebulae is a
result of enhanced dielectronic recombination in hot gas in the
nebular interior, perhaps driven by a hot stellar wind. 
  }
\listofauthors{D.~R.~Garnett \& H.~L.~Dinerstein}
\indexauthor{Garnett, D.~R.}
\indexauthor{Dinerstein, H.~L.}
\begin{document}

%% This command is necessary to typeset the title, abstract, etc. 
\maketitle

%%
%% And here starts the text....
%%
\section{Introduction}
\label{sec:intro}

Measurements of physical conditions and heavy-element abundances in 
photoionized nebulae rely heavily on bright, collisionally-excited
forbidden lines (FLs). However, abundances derived in this way are sensitive
to systematic errors in electron temperature and to deviations of
temperature from homogeneity because of the exponential temperature
dependence of forbidden-line emissivities, as pointed out by \scite{peim67}.

Recent observations of recombination lines (RLs) in a few nebulae
(\pcite{pst93}, \pcite{epte98}) have called into question the standard
forbidden line abundance analysis. These studies have found that
recombination lines give significantly higher abundances than the
forbidden lines. The common explanation is that the discrepancies
are due to temperature fluctuations, which cause the forbidden lines
to underestimate the true abundances. However, in some cases 
(e.g., \pcite{lsbc95}) the discrepancy between RL abundances and 
FL abundances is so large that the temperature fluctuation explanation
is difficult to accept. Therefore, it is necessary to study the 
recombination lines in more detail.

One problem has been that few nebulae have been studied in RL emission,
so that the systematics of RL emission variations are virtually unknown.
Here we report results from two new studies designed to examine RL
emission from a moderately-sized sample of PNe (\pcite{dlg00}, \pcite{gd00}),
in order to understand better the RL-FL abundance discrepancy.

\begin{figure}[!hb]
   \begin{center}
    \leavevmode
    \includegraphics[width=\columnwidth]{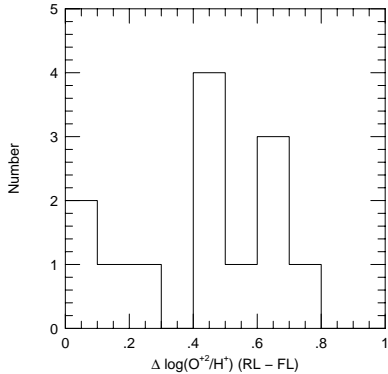}
    \caption{Histogram of the difference between O$^{+2}$ abundance
    derived from O~II and that derived from [O~III] in logarithmic
    units.  }
    \label{fig:histogram}
   \end{center}
\end{figure}
\section{New Observations of O II in PNe}
\label{sec:observations}

We observed O~II recombination lines in the 4100-4700 \AA\ region
of ten bright PNe with the Large Cassegrain Spectrograph on the 2.7m 
reflector at McDonald Observatory. We extracted 1-D spectra of these 
objects to study the O II lines at high signal/noise. The results are 
presented here. In addition, we obtained spatially-resolved long-slit
data for both O~II and [O~III] lines in eight PNe (six in common with 
the McDonald sample) using the B\&C spectrograph on the 2.4m Bok reflector
at Steward Observatory. These observations covered the 4150-5000 \AA\
region at 2 \AA\ resolution, similar to that of the McDonald spectra.
Here we present the results for one object, the Ring nebula NGC 6720.

\section{Results}
\label{sec:results}
  
Figure 1
%~\ref{fig:histogram} 
shows a histogram of the differences in O$^{+2}$ 
abundances 
derived from O II and [O III] in the McDonald PN sample. We
see a large spread in $\Delta$(O$^{+2}$/H$^+$), ranging from 0.0 to 0.8
dex. Thus, the RL-FL discrepancy is not seen in all nebulae.

\begin{figure}[!hb]
  \begin{center}
    \leavevmode
    \includegraphics[width=\columnwidth]{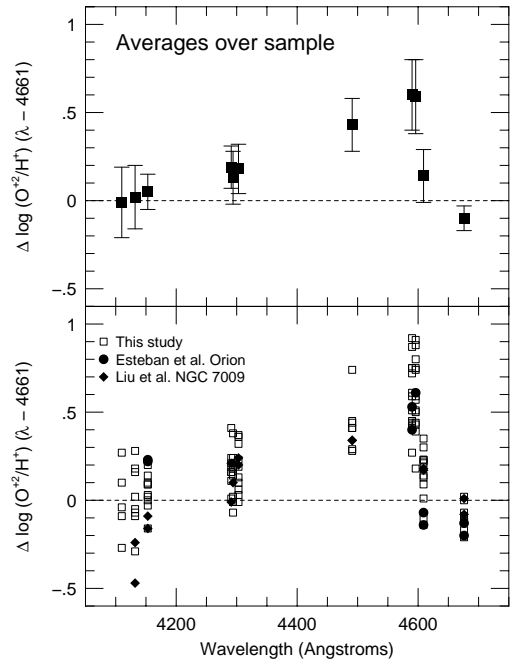}
    \caption{The logarithmic difference between O$^{+2}$ abundances from 
    O~II $\lambda$4661 and that derived from O~II lines in other multiplets. 
    The largest difference is seen for the multiplet 15 lines at 4590, 4596 \AA.}
    \label{fig:multiplets}
  \end{center}
\end{figure}
Next, we looked at the RL abundances by multiplet. The results are
shown in Figure 2,
%~\ref{fig:multiplets}
where we show the derived abundances 
for various
O II lines relative to that derived from O~II $\lambda$4661. 
The lower panel shows the individual results, while the upper panel
shows the differences averaged over the entire sample. The figure
shows that in most cases, the O~II lines give similar abundances. 
There are some notable exceptions,
however, in particular the lines from multiplet 15 at 4590, 4596 \AA.
The abundances from these lines are on average 0.6 dex greater than
those derived from the 4661, 4676 \AA\ lines of multiplet 1. The
3$p$ $^2$F$^o$ level from which multiplet 15 arises is populated
mainly by dielectronic recombination. We computed the abundances
using the dielectronic recombination coefficients of
\scite{ns84}, yet we see a large discrepancy with respect to
other O~II lines. This suggests that either the dielectronic 
recombination coefficients are underestimated, or that another
contribution to dielectronic recombination operates within the
affected PNe.

\begin{figure}[ht]
  \begin{center}
    \leavevmode
    \includegraphics[width=\columnwidth]{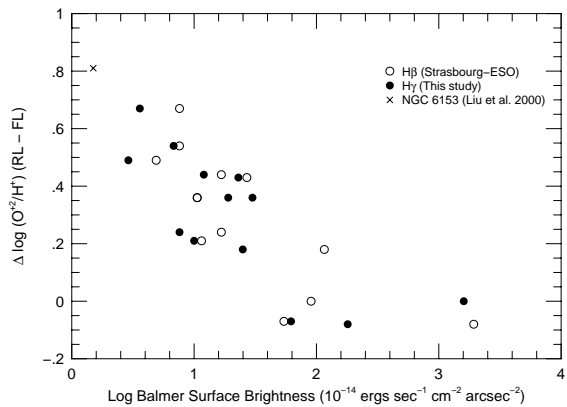}
    \caption{The difference between O$^{+2}$ abundances from O~II 
    and that derived from [O~III], plotted against nebular surface
    brightness. }
    \label{fig:surfbright}
  \end{center}
\end{figure}
Another intriguing correlation is shown in Figure 3, 
%~\ref{fig:surfbright}
where we show
the difference between RL and FL abundances plotted against the 
Balmer line surface brightness, which corresponds to the emission
measure. We see that the abundance discrepancy is tightly anti-correlated
with the nebular surface brightness. Similarly, we find that the
abundance discrepancy correlates with nebular diameter, such that the
largest PNe show the largest discrepancy between RL and FL abundances.
Compact, high-surface-brightness nebulae show little discrepancy between
RL and FL abundances. This suggests to us that the abundance problem
is a function of PN evolution, pointing to some physical process 
that is related to the evolutionary state of the nebula. 

\begin{figure}[ht]
  \begin{center}
    \leavevmode
    \includegraphics[width=\columnwidth]{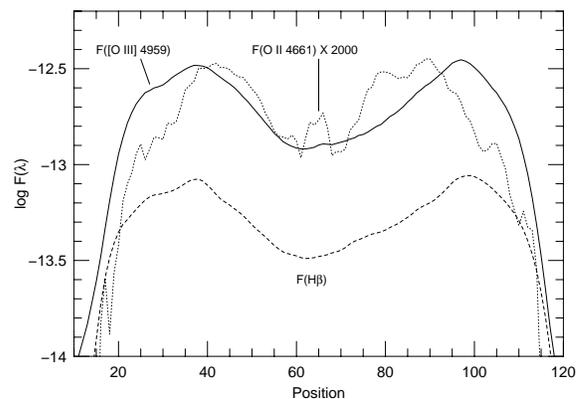}
    \caption{Log F($\lambda$) for [O~III] 4959 \AA\ ({\it solid line}), 
    H$\beta$({\it dashed line}), and O~II 4661 \AA\ ({\it dotted line})
    as a function of slit position across the Ring nebula. 
    The $\lambda$4661 fluxes have been multiplied by 2000 to aid the 
    comparison. The position is given in pixels, corresponding to 
    0\farcs86 per pixel. The central star is centered at pixel 66.
    }
    \label{fig:ringfluxes}
  \end{center}
\end{figure}
We now show some results from our spatially-resolved long-slit observations
of O~II and [O~III] in the Ring nebula NGC 6720. Figure 4
%~\ref{fig:ringfluxes}
shows the spatial distribution of the H$\beta$, [O~III] $\lambda$4959,
and O~II $\lambda$4661 fluxes along the slit. For reference, the slit
was aligned east-west and centered on the central star, which is located
at pixel 66 in the figure. We see from this plot that the O~II and [O~III]
emission are distributed very differently in the nebula. While the [O~III]
emission is distributed similarly to the H$\beta$ emission, the O~II
emission actually peaks $inside$ both [O~III] and H$\beta$. If the O~II
emission is produced mainly by radiative recombination, however, we would
expect the emission to peak $outside$ the [O~III] region, in the transition
from O$^{+2}$ to O$^+$. 
From these measurements we derived the O$^{+2}$/H$^+$ 
abundance ratio from both the RL and the FL along the slit. We see a large 
discrepancy between the RL abundance and the FL abundance in the central 
region of the nebula. The discrepancy decreases as one moves outward into 
the bright nebular shell. A similar result was found for the PN NGC 6153 
\cite{liu00}.

\section{Discussion and Speculation}
\label{sec:discuss}

%\begin{figure}
%  \begin{center}
%    \leavevmode
%    \includegraphics[width=\columnwidth]{GAD_f5.ps}
%    \caption{O$^{+2}$ abundances derived from O~II $\lambda$4661 and [O~III]
%    $\lambda$4959, plotted as a function of slit position. Reliable abundances
%    are obtained only for over the pixel range 20-110. }
%    \label{fig:ringabunds}
%  \end{center}
%\end{figure}
Our new observations provide hitherto unsuspected clues as to the cause of 
the discrepancy between RL and FL abundances. The discrepant abundances from 
multiplet 15 (Figure 2)
suggests that dielectronic recombination plays a larger role than 
previously suspected. The remarkable correlation with nebular surface
brightness in Figure 3 suggests that the cause of the RL-FL discrepancy
is related to the evolutionary state of the PN; larger, low average surface
brightness PNe show the largest differences between RL and FL abundances. 
The spatial distribution of the O~II 4661 \AA\ line in Figure 4 suggests
that some physical process in addition to radiative recombination contributes
to the O~II emission.

Can temperature fluctuations account for what we see? In order to force
the [O~III] lines to give the same abundances as O~II in the Ring nebula,
the mean $T_e$ in the central regions would have to be reduced from the
observed 12,000 K to 7,000 K, while increasing outwards to about 10,000 K
in the main shell. Such a thermal structure would be at odds with ionization
models for PNe with very hot ionizing central stars, unless the central
nebula is very metal-rich. However, we see no central enhancement of He, 
which might be expected in such a case. Similarly,
\scite{liu00} find that it is difficult to account for their observations
of NGC 6153 with the temperature fluctuation model. They find from new
observations made with $ISO$ that IR fine-structure lines in NGC 6153 
yield ion abundances similar to those obtained from the optical forbidden
lines, consistent with small temperature fluctuations. 
Liu et al. 2000
propose that the data can be accounted for by a model in which the RLs 
are emitted by dense, super-metal-rich clumps within the lower density
plasma where the FLs arise. 
While this model 
\adjustfinalcols
provides a natural explanation
and gives a good match to the data, it is not clear why such super-metal-rich
clumps should appear only in the most evolved nebulae, if our observed
correlation with surface brightness truly reflects nebular evolution.  

We propose a different model which could account for the observations.
We suggest that the RLs in PNe can be enhanced by high-temperature
dielectronic recombination to high $n$ levels that is not normally 
accounted for in 
nebular analysis. Planetary nebulae have been proposed to evolve
through the interaction of a fast central star wind with the slow
red giant wind of the progenitor. The fast central wind carves out
the central bubble and can generate hot gas. Dielectronic recombination
rates for various ions actually peak at log $T_e$ $\approx$ 5.0$\pm$0.5,
so hot gas at this temperature should show enhanced recombination line
emission. As the hot bubble expands, more of the nebula will show 
enhanced recombination line emission, which would account for the
correlation of the abundance discrepancy with nebular size, and for
the peaking of the RL emission interior to the [O~III] region. A test
of this model would be to look for a correlation between RL enhancements
and the presence of highly ionized species (e.g., O VI) marking the
presence of coronal gas.

Whatever the explanation for the discrepancy between RL and FL abundances,
a lot of new data will soon be available to test the various models. We
can look forward to a greatly improved understanding of this problem.

%% When using the rmaacite package, the \bibitem command should be of
%% the format: 
%%
%%             \bibitem[AUTHOR<YEAR>]{KEY} 
%%
%% so that the \cite{KEY}, etc. commands will work properly. 
%% 
%% If you are doing the citations manually, then you can just use
%% `\bibitem{}' instead. This will give you a warning about
%% `multiply-defined labels' which you can safely ignore.
%% 

\end{document}